\documentstyle[twoside,fleqn,espcrc2]{article}
\def\today{\ifcase\month\or
 January\or February\or March\or April\or May\or June\or
 July\or August\or September\or October\or November\or December\fi
 \space\number\day, \number\year}
\def\thebibliography#1{\section*{References\markboth
 {References}{References}}\list
 {[\arabic{enumi}]}{\settowidth\labelwidth{[#1]}
 \leftmargin\labelwidth
 \advance\leftmargin\labelsep
 \usecounter{enumi}}
 \def\newblock{\hskip .11em plus .33em minus .07em}
 \sloppy
 \sfcode`\.=1000\relax}

\def\r2{\sqrt 2}
\def\beq{\begin{equation}}
\def\eeq{\end{equation}}
\def\beqn{\begin{eqnarray}}
\def\eeqn{\end{eqnarray}}

\def\sinW2{\sin^2\theta_W}

\def\mz2{M_{z}^2}
\def\c2b{\cos 2\beta}

\def\mz{M_z}

\def\Fq2{F_{2}(q^2)}
\def\f{\({\cal F}\)}
\def\d1{{\f(\tilde c;\tilde s;\tilde W)+ \f(\tilde c;\tilde \mu;\tilde W)}}

\def\sec2w{sec^2\theta_W}

\title{Supergravity Unification}

\author{Ali H. Chamseddine\address{Center for Advanced Mathematical Sciences (CAMS) and\\
Physics Department, American University of Beirut, Lebanon}, R.
Arnowitt\address{Center for Theoretical Physics, Department of Physics, Texas A\&
M University, College Station TX 77843,-4242, USA} and Pran
Nath\address{Theoretical Physics Division, CERN CH-1211, Geneve 23, Switzerland}
\address{ Department of Physics, Northeastern University, Boston, MA 02115-5000,
USA}\thanks{Permanent address of P.N}}
\begin{document}

\begin{abstract}
A  review is given of the historical developments of 1982 that lead
to the supergravity unified model (SUGRA)
 with gravity mediated breaking of supersymmetry. Further
 developments and applications of the model in the period
 1982-85 are also discussed.  The supergravity unified model and  its
 minimal version
 (mSUGRA) are currently among the leading candidates for physics
 beyond the Standard Model. A brief note on the developments from
 the present vantage point is included.
\vspace{1pc}
\end{abstract}
\maketitle

\section{Introduction}

\indent

The main advantage of supersymmetry \cite{golfand} in the context of building
models of particle interactions is that it offers a technical solution to the
hierarchy problem of masses\cite{gildener}. Thus, for example,  the Higgs field
is not protected from acquiring large masses though loop corrections and in  a
grand unified theory the radiative corrections would give it a GUT size mass of
O($10^{14})$ GeV. However, if the scalar fields appear in  a supermultiplet then
supersymmetry would require a cancellation of the loop corrections in the
supersymmetric limit. When supersymmetry is broken the loop corrections to the
Higgs mass and to other masses will then have  a  dependence on the SUSY breaking
scale. So one finds that the breaking of supersymmetry is central  to getting  a
meaningful result out of supersymmetric theories.  Up to 1982 models in particle
physics
 were in the framework of global supersymmetry\cite{zumino}.
 However, this effort faced
an important hurdle, i.e., it is difficult to break
global supersymmetry in a phenomenologically acceptable fashion.
Thus efforts to break global supersymmetry spontaneously lead to
some unpleasant features as discussed below and consequently such efforts were
  abondoned.
An alternative suggestion, advanced however, was to introduce
in an ad hoc fashion soft breaking terms consisting of dimension
2 and dimension 3 operators in the theory which break supersymmetry
but nonetheless respect the ultraviolet behavior of the
theory and respect the gauge hierarchy since such terms
are super renormalizable\cite{dimo,gg}. The list of such soft terms consists of
 scalar mass terms such as for squarks and sleptons and for the
 Higgs, bilinear scalar terms,
  Majorana mass terms for the  gauginos, as well as
 cubic bosonic terms involving products of scalar fields.
However, such a
procedure introduces generally an enormous number of free parameters
in the theory. For the so called Minimal Supersymmetric Standard
Model this number is around 104. A theory such as this is not predictive.

    To understand the origin of soft breaking terms and
    to derive them from the basic model one needs to break
    supersymmetry spontaneously. However, the spontaneous
    breaking of supersymmetry leads to a Goldstone fermion,
    or a Goldstino and it is important to absorb this
    Goldstino to achieve a meaningful theory which is not
    in contradiction with experiment. To absorb the Goldstino
    one needs a vector-spinor,  i.e., a massless spin 3/2
    field or a gravitino. The necessity to get rid of
    the undesirable Goldstino requires that we consider
    an extension of supersymmetry to include gravity.
    In this extension the gravitino is the superparter of the
    graviton and thus one  is led to consider supergravity\cite{fnf}
    as the starting point  for a phenomenologically viable
    breaking of supersymmetry.
    However, to build models within supergravity
     one needs to couple  N=1 supergravity with
  chiral multiplets and a vector multiplet leading to the so
  called applied supergravity\cite{vonproyen,applied}.
  Remarkably the framework of applied
  supergravity allows one to break supersymmetry in a spontaneous
  fashion and also to build realistic models.
 One of the basic ingredients in the analysis
  is that the scalar potential of the theory is not positive
  definite which was one of the main hurdles in finding a global
  vacuum with broken supersymmetry in the framework of global
  supersymmetry. For the case of supergravity grand unification
  proposed in  1982\cite{chams}
  it is possible to break supersymmetry spontaneously in a so
  called  hidden sector and arrange the vacuum energy to vanish.
   The information of supersymmetry breaking from the hidden sector
   to the physical sector is communicated by gravity resulting in the
   growth of soft breaking terms in the
   physical sector\cite{chams,barbi}.
     For definiteness the
  GUT group chosen in the 1982 work was SU(5) although the main
  results of the effective theory are not tied to any specific
  GUT group\cite{chams,hlw,nac}.

    Supergravity grand unification model proposed in 1982
    also resolves another
    problem, i.e., the puzzle of the breaking of the electro-weak
    $SU(2)_L\times U(1)_Y$ symmetry. There is no natural
    mechanism for the breaking of the electro-weak symmetry in the
    Standard Model, and one arranges for the breaking by
    the introduction of a tachyonic mass term. However,
    the introduction of such a tachyonic mass term is rather
    ad hoc. Supergravity grand unification provides
    a natural mechanisms for the generation of the tachyonic
    mass term. In SUGRA unification one starts at the
    GUT scale with prescribed boundary conditions on the
    gauge coupling constants, Yukawa couplings and  on the
    soft parameters. The evolution of these parameters from
    the GUT scale down to the electro-weak scale turns the
    determinant of the Higgs mass square matrix negative
    which triggers the breaking of the electro-weak
    symmetry\cite{inou}.
    In the framework of MSSM there are two Higgs doublets in
    the theory, one which gives mass to the up quark and the
    other which gives mass to the down quark. These two doublets
    provide a bilinear term with a mixing parameter $\mu$ in
    the superpotential. Although this term is supersymmetric
    the $\mu$ parameter is of the size of the soft
    breaking terms. An important aspect of the radiative breaking
    of the electro-weak symmetry is that the spontaneous
    symmetry breaking conditions allow one to determine the
    $\mu$ parameter and this eliminates one of the
    arbitrary parameters in the theory.

    The radiative breaking of the electro-weak symmetry
    which arises in SUGRA model relates the weak scale
    $M_W$ with the soft SUSY breaking scale which in
    turn depends on the Planck constant.  The SUGRA
    model in this fashion connects two fundamental scales
    of physics, i.e., the Fermi scale and the Planck scale.
    Further, the fact that $M_W$ is determined in terms of soft
    parameters up to gauge and Yukawa coupling constant
    factors means that the soft parameters must lie in
    the TeV range, or else, one is faced with a new kind
    of a fine tuning problem. This result leads to the remarkable
    prediction in SUGRA GUTS that one must have sparticles
    lying in the TeV mass range.

    \section{SUGRA Unification}
    We discuss now the main features of the work of Ref.\cite{chams}.
    To build models containing gravity  one
    needs to couple supergravity with an arbitrary number of
     chiral multiples
$\lbrace\chi_i(x),z_i(x)\rbrace$ (where $\chi_i(x)$ are left (L) Weyl spinors and
$z_i(x)$ are scalar fields) and a vector multiplet which belongs to the adjoint
representation of the gauge group. Such a coupling scheme depends on three
arbitrary functions: :  the superpotential $g(z_i)$, the gauge kinetic function
$f_{\alpha\beta}(z_i$) (which enters in the Lagrangian as
$f_{\alpha\beta}F^{\alpha}_{\mu\nu}F^{\mu\nu\beta}$ with $\alpha,\beta$ = gauge
indices) and the  Kahler potential $d(z_i,z_i^{\dag}$).
 The bosonic part of the Lagrangian is given by

\begin{eqnarray}
L_B=-\frac{e}{2\kappa^2} R +\frac{e}{\kappa^2}G,^a_bD_{\alpha}z_a D^{\alpha}z^b
\nonumber\\ -\frac{e}{4}f_{\alpha\beta}F_{\mu\nu}^{\alpha}F^{\mu\nu\beta} -eV
\end{eqnarray}
Here $\kappa \equiv 1/M_{P\ell}$ where $M_{Planck}=2.4\times 10^{18}$ GeV,
G depends on the superpotential g(z),
and on the Kahler potential
$d(z,z^{\dagger})$ so that

\begin{equation}
G=-\frac{\kappa^2}{2}d(z,z^{\dagger}) -ln(\frac{\kappa^6}{4}|g|^2)
\end{equation}
and V is the
 effective potential for the scalar components of the chiral multiplets in
 the theory and is given by

\begin{equation}
V=-\frac{1}{\kappa^4}e^{-G}(3+G^{-1a}_bG_{,a}G,^{b})
+\frac{1}{8\kappa^4}(G,aT^{\alpha}z^a)^2
\end{equation}
As noted in the introduction the potential is no longer positive
definite because of the Planck scale corrections from supergravity.
Of course, in the limit
$M_{Planck}\rightarrow \infty$ one
recovers the result of global supersymmetry. However, the terms
proportional to $1/M_{PLanck}$  are essential
in generating an acceptable spontaneous breaking of supersymmetry.
The simplest way of breaking supersymmetry spontaneously is to
introduce a term  which is a chiral gauge
singlet and enters linearly in the superpotential, i.e.,
\begin{equation}
g_{SB}=m^2 (z+B)
\end{equation}
where the constant B is chosen to absorb the vanishing of the
cosmological constant.
This term generates a spontaneous breaking of supersymmetry
and gives a gravitino mass of size $m^2/M_{Planck}$.
However, it was later realized that the precise form of
the superpotential that breaks supersymmetry is not important,
and essentially what one needs is any form that breaks supersymmetry and
has the property that $g_{SB}=g_2(z)$ where\cite{nac}
\begin{equation}
g_{2}=\frac{m^2}{\kappa}f(\kappa z)
\end{equation}
so that at the minimum\cite{nac}

\begin{equation}
 <z>\sim \frac{1}{\kappa}, ~~~~<\kappa z>\sim O(1),  ~~~~
 <g_2>\simeq  \frac{m^2}{\kappa}
\end{equation}
and one has in this case that the gravitino mass is
\begin{equation}
m_{\frac{3}{2}}=\kappa^2<g_2>=\kappa m^2
\end{equation}

In addition to the bosonic interactions one also has interaction
structures in the theory which involve gauge couplings of the
gauginos to the chiral multiplets and these take the form

\begin{equation}
L_{\lambda}=-ig_{\alpha} \bar \lambda^{\alpha}z_i(\frac{T^{\alpha}}{2})_{ij}
\chi^j +h.c.
\end{equation}
 Further,
the superpotential contributes
to both the fermi and the bose interactions. Thus the Fermi interactions
for the Weyl spinors $\chi^i$ are given by\cite{salam}

\begin{equation}
L_g= -\frac{1}{2}(\bar \chi^{iC} g_{ij}\chi^j +h.c.)
\end{equation}

We now elaborate  on some of the further details in the construction
of SUGRA GUT models.
One of the immediate problems one faces in constructing such models
 is that of protecting the gauge hierarchy. Since
$<z>\sim O(M_{Planck})$
one finds that terms involving $z_iz_j z$ would develop masses which
are size $O(M_{Planck})$ and one has to suppress them.
A very
convenient way to overcome this problem is to use the separation
of the hidden and the physical sectors, i.e.,
place the chiral multiplets of the physical sector
in one part of the superpotential ($g_1$) and place the chiral fields
that break supersymmetry in another part of the superpotential
in the hidden sector ($g_2$) with no direct interaction between
the two sectors, i.e., one writes\cite{chams,nac}

\begin{equation}
g(z_i) = g_{1}(z_a) +g_{2}(z)
\end{equation}
where
$\lbrace z_i\rbrace=\lbrace z_a, z\rbrace$ where $z_a$ are physical
sector fields (squarks, sleptons, higgs) and z are the hidden sector fields
whose VEVs $\langle z\rangle ={\cal O}(M_{P\ell})$ break supersymmetry.
In this arrangement of separating the superpotential
into a physical sector
and a hidden sector one finds that the
 gauge hierarchy is
 maintained by the additive
nature of the physical and the hidden sector terms in the superpotential.
The actual proof of this remarkable theorem that the low
energy theory is protected  from corrections proportional to the
GUT scale and the Planck scale is rather subtle and we shall
discuss this shortly. For the  physical sector of the theory which
involves the quarks, the leptons and the higgs we take
the following potential

\begin{eqnarray}
g_1=\lambda_1 (\frac{1}{3}tr\Sigma^3 +\frac{1}{2}Mtr\Sigma^2) +\lambda_2
H'(\Sigma +3M')H\nonumber\\ +\lambda_3UH'H
+\epsilon_{xyzvw}H^xM^{yz}f_1M^{vw}\nonumber\\ +H'_xM^{xy}f_2M'_y
\end{eqnarray}
Here $M^{xy}$ and $M'_x$ are the 10 and $\bar 5$ of quarks and leptons,
$H'$ and $H$ are the $\bar 5$ and 5 of Higgs, U is a singlet
and $\Sigma$ is a 24 plet of SU(5). Minimization of the scalar
potential in this case gives the result

\begin{eqnarray}
 <\Sigma^x_y>=M Diag( 2,2,2,-3+\epsilon_2, -3-\epsilon_2)
\end{eqnarray}
and $<H^x>=<H'_x> \sim O(\kappa m^2)\delta^x_5$,
where $\epsilon_2$ depends on the soft SUSY parameters.
 Here one finds quite remarkably that supersymmetry breaking
 in the hidden sector  induces the breaking of $SU(2)_L\times U(1)_Y$
 so that $SU(2)_L\times U(1)_Y \rightarrow U(1)_{em}$. One notes
 that the breaking of $SU(2)_L\times U(1)_Y$ is semi-gravitational
 with $\kappa m^2\sim 300 $ GeV and can account for the W and Z mass.

 As noted earlier the miraculous thing in this model is that the
 integration of the heavy fields do not mix with the low energy
 sector of the theory. In general the presence of the large scale
 $M\sim M_G$ and $M_{Planck}$ introduces a new hierarchy
 problem as one can expect gravitaional
 corrections of sizes\cite{chams,nac}
 \begin{equation}
 (\kappa M)M, ~~(\kappa M)^2M,..,(\kappa M)^6M,..
 \end{equation}
 However, the detailed analyses show that all such corrections
 vanish, and  the low energy theory  contains only terms
 of $O(\kappa m^2)$.  An  analysis under the assumption of
 a flat Kahler potential and on
 integration over the heavy fields shows that the low energy
 theory can be parametrized by\cite{chams,hlw,nac}
\begin{equation}
V_{soft} =  m_0^2~z_az_a^{\dag}+\biggl[\frac{1}{3} { A}^{0}W^{(3)} +
\frac{1}{2}{B}^{0}W^{(2)}+h.c.\biggr]
\end{equation}
where $W^{(3)}$ is the cubic part and $W^{(2)}$ is the quadratic part of the
superpotential.
 Thus one finds that the soft SUSY breaking with the assumption
 of the flat Kahler potential generates a universal mass for the
scalar fields, and also  universal bilinear and trilinear couplings
arise in the theory. While  in the simple model with a
flat gauge kinetic energy function the gauginos are massless
at the tree level they do develop masses through loop corrections
where the GUT fields circulate in the loops with masses\cite{lajolla}
\begin{equation}
\tilde m_i =\frac{\alpha_i}{4\pi}m_{\frac{3}{2}}C \frac{D(R)}{D(A)}
\end{equation}
where $\alpha_i$ is the gauge coupling constant associated
with the sub groups $SU(3)$, $SU(2)$ and $U(1)$, and $C$
is the Casimir, D(R) is the dimensionality of the representation
exchanged, and D(A)  is the dimensionality of the
adjoint representation. One can also generate tree level
masses by assuming a field dependence in the gauge kinetic
energy function $f_{\alpha\beta}$. In this case one finds
that the gauginos gain  a mass at  $M_G$ of
($m_{1/2})_{\alpha\beta}\lambda^{\alpha}\gamma^0\lambda^{\beta}$
($\lambda^{\alpha}$ = gaugino Majorana field)  where

\begin{equation}
(m_{1/2})_{\alpha\beta}=\frac{1}{4} \kappa^{-1}\langle
G^i (d^{-1})^i_j~f_{\alpha\beta j}^{\dag}\rangle m_{3/2}
\end{equation}
\noindent
Here $G^i\equiv\partial G/\partial z_i^{\dag}$, ($d^{-1})^i_j$ is the matrix
universe of $d_j^i$ and $f_{\alpha\beta j} =\partial
f_{\alpha\beta}/\partial z_j$. For the case of the flat Kahler
potential where $d^i_j=\delta^i_j$ one finds that the gaugino
masses are universal at the GUT scale.  Further, more generally
one can write the gravitino mass so that
\begin{equation}
m_{\frac{3}{2}} =
\kappa^{-1}\langle exp [G/2]\rangle
 \end{equation}

    In the above we discussed how spontaneous supersymmetry
    breaking in the hidden sector induces soft
     breaking terms in the physical sector of the theory which
     in turn induce breaking of the electro-weak symmetry.
    The fact that the
  spontaneous breaking of the supersymmetry triggers the
  breaking of the electro-weak symmetry establishes  for the first
  time a direct connection between physics at the Planck scale and the
  physics at the electro-weak  scale.
     The electro-weak symmetry breaking in the model of Eq.(1)
     was induced
     at the tree level through an effective cubic interaction.
     Now it turns out that one can in fact do away with the cubic
     term and induce spontaneous breaking of the electro-weak
     symmetry via renormalization group effects.
     The basic idea here is that
  the one uses the GUT boundary conditions on the gauge coupling constants
  and the Yukawa couplings, and the boundary conditions
  on the soft parameters at the GUT scale and uses the renormalization
  group equations to evolve them down from the GUT scale to the
  electro-weak scale.
  The running of the renormalization group equations with at least
  one soft SUSY breaking parameter non-zero and the top quark Yukawa
  coupling turn the determinant of the Higgs mass matrix (and
  specifically the $H_2$  running $(mass)^2$) negative which triggers the
  breaking of the electro-weak symmetry.
  The minimization conditions for the scalar potential
  with respect to the Higgs VEVs
$v_{1,2} = \langle H_{1,2}\rangle$
  at the
  electro-weak scale provides the following two relations\cite{inou}
$\mu^2 ={(\mu_1^2-\mu_2^2tan^2\beta)}/{(tan^2\beta-1)} -
M_Z^2/2$; and $sin^2\beta$ =$ {(-2B\mu)}{(2\mu^2+\mu_1^2+\mu^2_2)}$.
Here $tan\beta = v_2/v_1$, B is the quadratic soft breaking parameter
($V_{soft}^B=B\mu H_1H_2$),  $\mu_i=m_{H_{i}}^2+\Sigma_i$, and
$\Sigma_i$ are loop corrections.
In the above all parameters are running parameters at the
electroweak scale $Q$ which can be taken to be $Q=M_Z$. The first
relation above
determines $\mu$ while the second  allows one to
eliminate B in favor of $tan\beta$.
This determination increases  the predictivity of the model.
With inclusion of the constraints of the electro-weak symmetry
breaking the minimal SUGRA model (mSUGRA) depends on four
soft breaking parameters and one sign
\begin{equation}
m_0, m_{1/2}, A_o, B_0, sign(\mu)
\end{equation}
\noindent
Alternately one may choose
$m_0$,  $m_{\tilde g}$, $A_t$, $tan\beta$, and sign($\mu$)
 as the independent parameters.  If one allows for complex
 soft parameters then after field redefinitions there are two
 additional parameters which can be chosen to be the phase
 of $A_0$ and the phase of $\mu$. The appearance of the
 phases brings into the theory in a natural way new sources
 of CP violation over and above the CP violation in the CKM
 mass matrix. Further, the nature of physics at the Planck scale
 is not fully understood and it is likely that the assumption of
 a flat Kahler potential should be relaxed by the introduction of
 curvature terms. Such curvature terms sometimes appear in string models.
 The appearance of non-flatness in the Kahler potential generates
 non-universalities in the soft parameters enlarging  the number of
 parameters in the theory\cite{soni}.
 Further, the picture
     of supersymmetry breaking via chiral scalar terms in the
     hidden sector of the theory is an effective treatment of
     some more basic phenomenon in the underlying fundamental
     theory which is still unknown. Several possibilities for
     the nature of such a phenomena including the possibility
     of condensates breaking supersymmetry have been
     discussed\cite{nilles} in the literature.
  In addition, most string models that accomodate grand unification
  at the GUT scale reduce to a SUGRA model below $M_{GUT}$, and thus
  mSUGRA remains  a bench mark for testing experimental
 data with theory.  Next we discuss the sparticle
 spectrum that emerges in such a theory at low energy.

  The minimal SUGRA model with the soft SUSY parameters defined
  by Eq.(18) generates a low energy mass spectrum for the Higgs
  and for the sparticles which possesses identifiable properties.
  The Higgs sector of the theory gives three neutral fields,
  one of which is  a pseudo-scalar (A) and the other two are scalars
  (h,H).
  The masses of the scalars are related to the masses of the
  pseudo-scalars by the relation\cite{tata}

\begin{eqnarray}
  m_{h,H}^2=\frac{1}{2}[M_Z^2+M_A^2\pm ((M_Z^2+M_A^2)^2\nonumber\\
  -4(\cos 2\beta)^2
  m_Z^2M_A^2)^{\frac{1}{2}}]
  \end{eqnarray}
 In addition there is a charged Higgs boson whose mass is
 given by
 \begin{equation}
 M_{\pm}^2=M_W^2+M_A^2
 \end{equation}
 The mass relations for the scalar Higgs lead  to the result\cite{tata}
 \begin{equation}
 m_h\leq M_Z
 \end{equation}
 Of course, this is a relation at the tree level and there are
 important loop corrections to this formula.

 One can also discuss the masses and the couplings
 for the sparticles that appear
 in the theory.
 The superpartners in the theory consist of gauginos, Higgsinos
 and sfermions. The gaugino-Higgsino sector involves
 the trilinear Higgs-Higgsino-gaugino coupling which  after
 spontaneous breaking of the electro-weak
  symmetry generates a gaugino -Higgsino mixing term.
In the
charged gaugino-Higgsino sector one finds  a mass matrix
(in the $\tilde W^{\pm}-\tilde H^{\pm}$ basis) of the form\cite{model}

\medskip
\begin{equation}
M_{\chi}^{\pm}=\left(\begin{array}{cc}
{\tilde m}_2 & \sqrt 2 M_W sin\beta\\
\\
\sqrt 2M_W cos\beta &\mu
\end{array}\right)
\end{equation}
The eigen-vectors of this matrix (Winos) have the mass eigen-values

\begin{equation}
\tilde m_{\pm}=\frac{1}{2}|[4\nu_+^2+ (\mu -\tilde m_2)^2]^{\frac{1}{2}}
\pm [4\nu_-^2+(\mu +\tilde m_2)^2]^{\frac{1}{2}}|
\end{equation}
where $\tilde m_{\pm}=|\lambda_{\pm}|$ and $\lambda_{\pm}$ are
the eigen-values of the M, and $\nu_{\pm}$ are defined by

\begin{equation}
\sqrt 2 \nu_{\pm}=M_W (sin\beta \pm \cos\beta)
\end{equation}

\medskip
In the neutral gaugino - Higgsino sector one finds
 (in the ($\tilde W_3,~\tilde B,~\tilde H_1,~\tilde H_2$) basis)
 the mass matrix
\medskip
\begin{equation}
M_{\chi}^0=\left(\begin{array}{cccc}
\tilde m_2 & o & a & b\\
\\
o & \tilde m_1 & c & d\\
\\
a & c & o & -\mu\\
\\
b & d & -\mu & o
\end{array}\right)
\end{equation}

\medskip
\noindent
where $a=M_Z cos\theta_W cos\beta$, $c=-tan\theta_Wa$,
$b=-M_Zcos\theta_W$ sin$\beta$,  d=-tan$\theta_Wb$ and $\theta_W$ is the weak
mixing angle. The eigen-vectors of this mass matrix (Zinos) have a
somewhat complicated analytical form since the eigen value equation
here is a quartic one.
 However, the structure  of the matrix reveals that there
are regions where some  Zinos may be mostly either gauginos or
 Higgsinos.

For the  sfermions it is found that the masses receive contributions
from the D terms  and from the gaugino loop corrections so
that at low energy one has, example for $\tilde u_L$\cite{model,inou},

\begin{eqnarray}
m_{\tilde u_{L}}^2 = m_0^2 + m_u^2 +\tilde \alpha_G[\frac{8}{3}f_3+\frac{3}{2}
f_2+\frac{1}{30}f_1] m_{\frac{1}{2}}^2\nonumber\\ + \left({1\over 2}-{2\over 3}
sin^2 \theta_W\right) M_Z^2 cos 2\beta
\end{eqnarray}
and for $\tilde u_R$ one has
\begin{eqnarray}
m_{\tilde u_{R}}^2 = m_0 + m_u^2 + \tilde \alpha_G[\frac{8}{3}f_3+\frac{8}{15}f_1]
  m_{\frac{1}{2}}^2\nonumber\\ +{2\over 3} sin^2\theta_W M_Z^2 cos 2\beta
\end{eqnarray}
Here $f_i(t)=(1-1/(1+\beta_it)^2)/\beta_i$ (i=1,2,3) where
 $t=ln(M_G^2/Q^2)$, $\beta_i=(b_i/4\pi)\tilde\alpha_i$, where
 $b_i=(33/5,1,-3)$ for U(1), SU(2) and SU(3), and
 $\tilde\alpha_G=\alpha_G/(4\pi)$,
where $f_i$ are the form factors. Similar relations hold for
the sleptons. Using the above analysis one can
  write the low energy effective Lagangian in terms of the
diagonalized fields. This effective Lagrangian was given in
Ref.\cite{model,applied,tata}.

Soon after the invention of SUGRA model there was an intense
activity to elucidate its experimental implications. Since the
values of $m_0$ and $m_{\frac{1}{2}}$ were not fixed by the theory
(and remain still to be fixed by experiment) it was natural
then to assume the lowest values compatible with experiment at
that time. One of the important signals for SUGRA models with
R parity invariance that emerged was the missing energy signal.
Thus supersymmetric decays of the W and Z were widely discussed
 such as $W\rightarrow \tilde W+ \tilde Z$ and
 $Z\rightarrow \tilde W +\tilde Z$ as well as
decays of the Wino, the neutralino and the sfermions, etc
were widely studied\cite{swein,signal}
\begin{eqnarray}
\tilde W\rightarrow l+\bar \nu_l+ \tilde Z_i,
,q_1+\bar q_2+\tilde Z_i,~~h+\tilde Z_i,..\nonumber\\
\tilde Z_k\rightarrow ^{\pm}+\tilde W^{\mp}, ~~l+\bar l+\tilde Z_j,..\nonumber\\
\tilde f\rightarrow f+ \tilde Z,~~f_1+\tilde W,..
\end{eqnarray}
In all these decays the identifying signal is large missing
energy associated with the disappearance of the lightest
neutral particle which is absolutely stable under the
assumption of R parity invariance.
Missing energy signals at colliders also arise, such as
in sparticle ($\tilde s$) pair production
$p\bar p\rightarrow \tilde s+\tilde s +X$.
Each sparticle is expected to decay producing
missing energy.
An important implication of
the fact that the lightest supersymmetric particle (lsp) would be
absolutely stable under R parity conservation is that such
a particle is a candidate for non-baryonic cold dark matter.
Specifically, the lightest neutralino appears a good  candidate
for the LSP\cite{goldberg}.

Another important implication of SUGRA model that was analysed
was the supersymmetric correction to the muon anomalous moment
$g_{\mu}-2$.
    Thus soon after the formulation of the SUGRA models analyses
    of $g_{\mu}-2$ were carried out using the SUGRA unification
    framework\cite{yuan}. These represented the first
    realistic and reliable analyses of the effects of
    supersymmetry on the muon anomaly. Since SUGRA provided
    the first realistic framework  it was now possible to
    make quantitative predictions. Specifically it
    was pointed out that the effect of the electro-weak
    corrections could be {\it as large or larger} than the
    Standard Model electro-weak correction
    and any experiment which tests the
    SM electro-weak contribution to $g_{\mu}-2$
    would also test the supersymmetric correction.
    This result played a role in the pursuit to
    improve the experimental limits on $g_{\mu}-2$
    and provided another reason aside from checking the
    Standard Model result to get a better measurement
    when the Brookhaven experiment E821 was being
    proposed around 1984.

    We turn now to another aspect
    of SUGRA models which pertains to baryon and lepton number
    violation in the model. Here on must distinguish between
    two aspects to SUGRA unification, the first being a specific mechanism
    for the spontaneous breaking of supersymmetry in the
    hidden sector and its transmittal to the visible
    sector and the second being the GUT structure of the
    theory. The first aspect of the theory, i.e., growth of
    soft breaking terms protects baryon and lepton number
    conservation under the constraint of R parity invariance.
    Further, under the constraint of R parity invariance
    baryon and lepton number violating dimension four operators
    are also absent which eliminates rapid proton decay.
    However, even under the constraint of R  parity one can
    generate dimension five operators which violate baryon
    and lepton number and lead to nucleon instability\cite{wein}.
    However, the nature of the dimension five operator
    depends on the nature of the GUT group and there are
    several possibilities here such as SU(5),
    SO(10), E(6) etc. Much of the early work in proton
    stability was in the framework of the minimal SU(5)
    model since such a group also appears in the breaking
    of larger GUT groups. However, what is unique about
    proton decay via dimension  five operators is that
    it depends on both the GUT structure which generates
    the baryon and lepton number violation in the theory
    and on the soft breaking sector of the theory which
    enters in the dressing of the dimension five operators
    and leads to dimension six operators after dressing
    with the chiral
    structure LLLL, LLRR, RRLL and RRRR for the four
    fermi interaction that governs proton decay.
     Thus after the formulation of
    the SUGRA models it was possible to carry out detailed
    analyses of proton decay amplitudes using the framework
    of SUGRA unification in the dressing loop diagrams.
    This analysis was carried out in
    Refs.\cite{acn,nac1}. As is evident predictions of the proton
    lifetime
    depend on the low energy parameters and on the
    GUT parameters. If one has sufficient information on the
    low energy parameters then that would lead to a constraint on
    the nature of GUT physics.\\

{\bf 6.~~A View from the Present}\\
    In the intervening period since 1982-85 -present
     there has been considerable further activity in
    the applications of SUGRA models. Further, there
    is now beginning to accumulate some indirect evidence
    that generally favors SUGRA unification.
    Recall that in Sec.2 one discussed a crucial aspect of
    SUGRA unification which is the connection between the
    soft breaking scale and the electro-weak scale. Because
    of this connection one finds that the sparticle masses should
    be in the TeV range.  This result finds an indirect support
    in the high precision LEP data. Thus the LEP data measures
     the gauge coupling constants $\alpha_1, \alpha_2$ and
     $\alpha_3$ to a great precision. Extrapolation of these
     coupling constants to high scales shows that they do not
     meet at a point using the Standard Model spectrum
     but do meet in MSSM with the supersymmetric particle
     spectrum with sparticle masses typically in the TeV
     range. The crucial ingredient in the unification of
     the gauge couplings is the fact that the SUSY particles
     with masses in the TeV range give just the right amount
     of corrections to the beta functions for all the three
     couplings to meet at a point. Thus one of the predictions
     of SUGRA GUTs that sparticle masses must lie in the
     TeV range finds support in the high precision LEP data.
    The theoretical analyses using SUGRA unification produce
    a unification of the gauge couplings to within 2 sigma.
    However, there could be additional contributions due to
    Planck scale corrections, which induce terms via
    corrections to the gauge kinetic energy term.

    Since the early work of 1982-1985 the experimental lower limits
    on the sparticles have moved up eliminating many of the possible
    decay channels such as the supersymmetric decays of the
    W and Z. However, the off shell decay of particles  such
    as the decay
    $W^*\rightarrow\tilde W+\tilde Z_2\rightarrow l_1+l_1+l_2+ E_T$
    produces a trileptonic signal which is one of the prime signals for
    the discovery of Winos in the Tevatron RUNII and at the LHC.
    Further, there is now a considerable body of work which
    analyses the signals in the SUGRA models in an array
    of channels depending on the nature of the sparticles produced
    (see, eg., SUGRA Working Group Collaboration, hep-ph/0003154).
    A very encouraging sign for weak scale supersymmetry is
    the recent precise determination of the muon g-2 at Brookhaven
    (H.N. Brown et.al., Muon (g-2) Collaboration,
    hep-ex/0102017).
    The experiment finds a difference at the 2.6 sigma level
     between the experimental value of $a_{\mu}$
    (where $a_{\mu}=(g_{\mu}-2)/2$) and its prediction in
    the Standard Model so that $a_{\mu}^{exp}-a_{\mu}^{SM}$=
    $43(16)\times 10^{-10}$. This result indicates a very significant
     correction
    to the Standard Model from new physics. Now the prime
    candidate for this new physics is supersymmetry and specifically
    we note at this point that in the well motivated SUGRA model
    the supersymmetric electro-weak correction was predicted to
    be of the size of the Standard Model electro-weak correction
    or larger\cite{yuan}. Further, analysis of this correction within
    mSUGRA shows that the Brookhaven result implies the existence
    of some low lying particles which can be produced
    at the LHC and possibly in RUNII of the Tevatron. Test of
    this prediction  await experiments at RUNII and at the
    LHC.

  {\bf 6.~~ Conclusion}\\
  Supergravity unification proposed 19 years ago remains
  one of the leading candidates for physics beyond the Standard Model.
  The model is consistent with all known data and predicts
  new physics within reach of accelerators, specifically the LHC and
  also possibly within reach of RUNII of the Tevatron.
  Deviations from the Standard Model are determined by the soft breaking
  sector of the theory which is governed by Planck scale physics.
  Thus observation of deviations from the Standard model will provide
  information about the nature of physics at the Planck scale
  and possibly about the  nature of the underlying string model.

\noindent
{\bf Acknowledgemnets:} This work was supported in part by
NSF grant numbers  PHY-0070964 and PHY-9901057.\\

\end{document}